\documentstyle[pra,aps,psfig]{revtex}
\def\Im{\mathop{\rm Im}}

\begin{document}
\title{Inertia of Casimir energy}
\author{Marc Thierry Jaekel $^{(a)}$ and Serge Reynaud $^{(b)}$}
\address{(a) Laboratoire de Physique Th\'{e}orique de l'ENS
\thanks{%
Unit\'e propre du Centre National de la Recherche Scientifique,
associ\'ee \`a l'Ecole Normale Sup\'erieure et \`a l'Universit\'e
Paris-Sud}, 24 rue Lhomond, F75231 Paris Cedex 05 France\\
(b) Laboratoire de Spectroscopie Hertzienne
\thanks{%
Unit\'e de l'Ecole Normale Sup\'erieure et de l'Universit\'e
Pierre et Marie Curie, associ\'ee au Centre National de la Recherche
Scientifique}, 4 place Jussieu, case 74, F75252 Paris Cedex 05 France}
\date{{\sc Journal de Physique I} {\bf 3} (1993) 1093-1104}
\maketitle

\begin{abstract}
Moving mirrors are submitted to reaction forces by vacuum fields. The
motional force is known to vanish for a single mirror uniformly accelerating
in vacuum. We show that inertial forces (proportional to accelerations)
arise in the presence of a second scatterer, exhibiting properties expected
for a relative inertia: the mass corrections depend upon the distance
between the mirrors, and each mirror experiences a force proportional to the
acceleration of the other one. When the two mirrors move with the same
acceleration, the mass correction obtained for the cavity represents the
contribution to inertia of Casimir energy. Accounting for the fact that the
cavity moves as a stressed rigid body, it turns out that this contribution
fits Einstein's law of inertia of energy.

\end{abstract}

Scatterers in vacuum are submitted to the radiation pressure of vacuum
fields. In a configuration with two motionless mirrors, a mean force, the
so-called Casimir force, results for each of them \cite{Inertia1,Inertia2}.
As known since Einstein \cite{Inertia3}, the field energy stocked inside a
box contributes to its inertia. This law of inertia of energy has to be
valid for any kind of energy bound to the motion of any system \cite
{Inertia4}. Casimir energy corresponds to the particular case of a
Fabry-Perot cavity, a system formed by two mirrors which models Einstein's
``box'', immersed in vacuum fields, and can be considered as a small amount
of vacuum energy stocked inside the cavity, actually a negative amount as a
binding energy for an atomic system. As a consequence of the law of inertia
of energy, the inertial mass of the cavity has to vary with Casimir energy
for example, when the length is varied.

Because of the infiniteness of the vacuum energy density (when integrated
over frequency) however, it has often been claimed that vacuum energy is not
a real energy like photons. Particularly, it seems to be admitted that it
does not gravitate \cite{Inertia5} so that its contribution to inertial
forces can also be questionned. Nevertheless, it has also been argued that
Casimir energy, a finite energy difference between two vacuum
configurations, has to contribute to gravitation and inertia \cite{Inertia6}. 
In the present paper, we demonstrate that Casimir energy does indeed
contribute to inertia of the Fabry-Perot cavity. For this demonstration, we
use previously obtained results providing the motional forces, i.e. the mean
forces experienced by mirrors moving in vacuum \cite{Inertia7,Inertia8}
which are associated with the quantum fluctuations of vacuum radiation
pressure \cite{Inertia9}.

For a perfectly reflecting mirror alone in the vacuum state of a scalar
field in a two dimensional (2D) spacetime \cite{Inertia10}, the motional
force $\delta F$ can be written in a linear approximation in the
displacement $\delta q$ as 
\begin{equation}
\delta F(t)=\frac{\hbar }{6\pi c^ 2 }\delta q^{\prime \prime \prime }(t) 
\eqnum 1 
\end{equation}
This force vanishes for a uniform velocity, as well as for a uniform
acceleration, which can be interpreted as a consequence of spatial
symmetries of the vacuum. Vacuum fields are invariant under the action of
Lorentz boosts \cite{Inertia11}, so that the motional force vanishes for
uniform velocity. When seen by a uniformly accelerating observer, vacuum
fields appear as thermal fields for a motionless observer \cite{Inertia12}
and the motional force vanishes for uniform acceleration. These properties
remain true for a partially transmitting mirror in vacuum \cite
{Inertia7,Inertia13}.

The motional forces have also been computed in the configuration of a
Fabry-Perot cavity in vacuum \cite{Inertia10,Inertia8}, where the spatial
symmetries previously discussed are broken. We show in the present paper
that they contain inertial forces, proportional to the accelerations, which
exhibit properties expected for a relative inertia: the mass corrections
depend upon the distance, and forces are obtained for each mirror, which are
proportional to the acceleration of the other one. We will also compute the
mass correction for a global motion of the system. In the limiting case of
perfect mirrors for instance, we obtain a mass correction which is twice the
value of Casimir energy over $c^ 2$. Taking into consideration that the
cavity moves as a stressed rigid body (same motion for the two mirrors), a
situation elucidated by Einstein himself \cite{Inertia14}, it turns out that
this is exactly the prediction of the law of inertia of energy.

\section*{Perfect mirrors: qualitative derivation of inertia corrections}

We first study the situation of two perfect point-like mirrors in the vacuum
state of a 2D scalar field. All relations, including the definition of
vacuum, will refer to an inertial frame.

A linear approximation of the expression obtained in this case by Fulling
and Davies \cite{Inertia10} provides us with the force $\delta F_1$
exerted upon the mirror 1 as a function of the positions $\delta q_1$ and 
$\delta q_2$ of the two mirrors \cite{Inertia15}; more precisely, $\delta
F_1$ is the response of the mean force to classical displacements) 
\begin{eqnarray}
\delta F_1 (t) &=&\frac{\hbar }{6\pi c^ 2 }\left( \delta q_1 ^{\prime
\prime \prime }(t)-\delta q_2 ^{\prime \prime \prime }(t-\tau )+\delta
q_1 ^{\prime \prime \prime }(t-2\tau )-\delta q_2 ^{\prime \prime \prime
}(t-3\tau )+\ldots \right)   \nonumber \\
&&+\frac{\hbar \pi }{6c^ 2 \tau ^ 2 }\left( \frac 1  2 \delta q_1 ^{\prime
}(t)-\delta q_2 ^\prime (t-\tau )+\delta q_1 ^\prime (t-2\tau )-\delta
q_2 ^\prime (t-3\tau )+\ldots \right)   \eqnum 2 
\end{eqnarray}
where $\tau $ is the propagation delay of light from one mirror to the other
and $q$ the distance between the mirrors 
\[
\tau =\frac{q}{c}
\]

The terms proportional to third order derivatives in equation (2) have the
same form as the damping force (1) for a single mirror, but the modification
of the stress tensor generated by the mirrors' motion now propagates from
one mirror to the other and is reflected back by the mirrors. This is why
the time of flight $\tau $ appears in the expression of the force. Although
they have the same form as for a single mirror, these terms give rise to
mass corrections. A qualitative demonstration goes as follows. Extracting
the contribution of these terms to the motional force (first line of eq. 2)
and considering the quasistatic limit where the positions vary slowly on a
time scale $\tau $, one transforms the sum over discrete times into an
integral 
\[
\delta F_1 (t)\approx {\int^{t}}{\rm d}t^\prime \frac{\hbar }{12\pi
c^ 2 \tau }\left( \delta q_1 ^{\prime \prime \prime }(t^\prime )-\delta
q_2 ^{\prime \prime \prime }(t^\prime )\right) 
\]
and one gets a motional force depending upon the accelerations of the two
mirrors 
\[
\delta F_1 (t)\approx \frac{\hbar }{12\pi cq}\left( \delta q_1 ^{\prime
\prime }(t)-\delta q_2 ^{\prime \prime }(t)\right) 
\]

The other terms of equation (2), proportional to velocities, are not present
in the one mirror problem. They are associated with the existence of a
static Casimir force, since their contribution (second line of eq. 2) leads
in the quasistatic approximation to 
\[
\delta F_1 (t)\approx \frac{\hbar c\pi }{12q^{3}}\left( \delta
q_1 (t)-\delta q_2 (t)\right) 
\]
This is the variation with the distance $q=q_2 -q_1 $ of the mean Casimir
force $F_1 $ 
\begin{equation}
F_1 =\frac{\hbar c\pi }{24q^ 2 }=\partial _{q}U\qquad U=-\frac{\hbar c\pi }
{24q}  \eqnum{3}
\end{equation}
where $U$ is the known expression for the Casimir energy \cite{Inertia2}.

In the approximate expressions obtained in this section, the position or
acceleration of one mirror are measured relatively to the position or
acceleration of the other one. In the more precise discussion which follows,
this property will remain true for the positions (the static force only
depends upon the distance between the two mirrors), but accelerations will
appear only partly as relative quantities; this feature will lead to a non
vanishing correction for the global mass of the cavity.

\section*{Perfect mirrors: quantitative evaluation of inertia corrections}

The motional forces $\delta F_{i}$ can be written in the temporal or
spectral domains 
\begin{eqnarray*}
&&\delta F_{i}(t)={\int }{\rm d}\tau \sum_{j}\chi _{ij}(\tau )\delta
q_{j}(t-\tau ) \\
&&\delta F_{i}[\omega ]=\sum_{j}\chi _{ij}[\omega ]\delta q_{j}[\omega ]
\end{eqnarray*}
where we denote for any function $f$ 
\[
f(t)=\int \frac{{\rm d}\omega }{2\pi }f[\omega ]e^{-i\omega t}
\]

The expressions (2) for the motional forces correspond to the following
susceptibility functions where the real and imaginary parts are separated in
the spectral domain 
\begin{eqnarray}
\chi _{ij}[\omega ] &=&\chi _{ji}[\omega ]=\widetilde{\xi }_{ij}[\omega
]+i\xi _{ij}[\omega ]  \eqnum{4a} \\
\xi _{11}[\omega ] &=&\xi _{22}[\omega ]=\frac{\hbar }{12\pi c^ 2 }\omega
^{3}  \eqnum{4b} \\
\xi _{12}[\omega ] &=&0  \eqnum{4c} \\
\widetilde{\xi }_{11}[\omega ] &=&\widetilde{\xi }_{22}[\omega ]=-\frac
\hbar {12\pi c^ 2 }\frac{\omega ^{3}-\omega \frac{\pi ^ 2 }{\tau ^ 2 }}
{\tan (\omega \tau )}  \eqnum{4d} \\
\widetilde{\xi }_{12}[\omega ] &=&\frac{\hbar }{12\pi c^ 2 }\frac{\omega
^{3}-\omega \frac{\pi ^ 2 }{\tau ^ 2 }}{\sin (\omega \tau )}  \eqnum{4e}
\end{eqnarray}
It can be noted that the dissipative parts $\xi _{ij}$, imaginary parts of 
$\chi _{ij}$ and odd functions of $\omega $, coincide with the contributions
of the outer space (for each mirror, only one half of the outer space
contributes and $\xi _{ii}$ is half the value of $\xi $ for a single mirror)
while the dispersive parts $\widetilde{\xi }_{ij}$, real parts of $\chi _{ij}
$ and even functions of $\omega $, are the contributions of the intracavity
space. This fact has a clear interpretation: the outer fields constitute an
open quantum system, corresponding to a continuous spectrum; in contrast,
the intracavity fields are characterized by a discrete spectrum and are
unable to contribute to dissipation. As a consequence, there is no
dissipative part in the mutual susceptibility ($\xi _{12}=0$).

The dissipative parts $\xi _{ij}$ are the commutators of the force operators
and can be deduced from the correlation function $C_{ij}$ computed for
motionless mirrors \cite{Inertia7,Inertia8} 
\[
\xi _{ij}(t)=\frac{\left\langle \left[ F_{i}(t),F_{j}(0)\right]
\right\rangle }{2\hbar }=\frac{C_{ij}(t)-C_{ji}(-t)}{2\hbar }\qquad
C_{ij}(t)=\left\langle F_{i}(t)F_{j}(0)\right\rangle -\left\langle
F_{i}\right\rangle \left\langle F_{j}\right\rangle 
\]
Fluctuations can also be recovered from dissipation through the relation 
\cite{Inertia7,Inertia8} 
\[
C_{ij}[\omega ]=2\hbar \theta (\omega )\xi _{ij}[\omega ]
\]
i.e. the fluctuation-dissipation relation \cite{Inertia16} at the limit of
zero temperature.

The dispersive functions $\widetilde{\xi }_{ij}$ diverge at the zeros 
$\omega =m\frac{\pi }{\tau }$ of the denominators, except for $m=0$ or $m=1$
where the numerators vanish. These divergences result from a constructive
interference between the different numbers of cavity roundtrips \cite
{Inertia8}. According to causality (which is apparent in eqs 2), the
susceptibility functions (4) are analytic in the upper half-plane of the
frequency domain ($\Im \omega >0$), and the dispersive parts are related
to the dissipative ones through dispersion relations, a property which is
more easily checked for partially transmitting mirrors (see the next
section).

As the susceptibility functions are regular around $\omega =0$, a
quasistatic expansion of the force may be performed, in which coefficients
are introduced for describing static, viscous and inertial forces (we will
not be interested in higher order quasistatic coefficients) 
\begin{eqnarray}
&&\delta F_{i}(t)=-\sum_{j}\left( \kappa _{ij}\delta q_{j}(t)+\lambda
_{ij}\delta q_{j}^\prime (t)+\mu _{ij}\delta q_{j}^{\prime \prime
}(t)+\ldots \right)   \eqnum{5a} \\
&&\chi _{ij}[\omega ]=-\kappa _{ij}+i\omega \lambda _{ij}+\omega ^ 2 \mu
_{ij}+\ldots   \eqnum{5b} \\
&&\kappa _{ij}=-\chi _{ij}[0]\qquad \lambda _{ij}=-i\chi _{ij}^{\prime
}[0]\qquad \mu _{ij}=\frac{\chi _{ij}^{\prime \prime }[0]} 2   \eqnum{5c}
\end{eqnarray}
The quasistatic coefficients are deduced from equations (4); the static
coefficients $\kappa _{ij}$ describe the variation with the distance $q$ of
the force $F=F_1 =-F_2 $ 
\begin{equation}
\kappa _{11}=\kappa _{22}=-\kappa _{12}=-\kappa _{21}=-\frac{\hbar c\pi }
{12q^{3}}=\frac{{\rm d}F}{{\rm d}q}  \eqnum{6a}
\end{equation}
The viscosity coefficients $\lambda _{ij}$ vanish while the inertia
corrections $\mu _{ij}$ are given by 
\begin{eqnarray}
\mu _{11} &=&\mu _{22}=-\frac{\hbar }{12\pi cq}\left( 1+\frac{\pi ^ 2 }{3}
\right)   \eqnum{6b} \\
\mu _{12} &=&\mu _{21}=-\frac{\hbar }{12\pi cq}\left( -1+\frac{\pi ^ 2 }{6}
\right)   \eqnum{6c}
\end{eqnarray}
The mass corrections differ from the approximate ones previously discussed
by numerical factors: the terms in equations (2) which are proportional to
velocities also contribute to the inertial corrections; $\mu _{11}$ and $\mu
_{12}$ no longer have opposite values and they actually have the same sign.
This will allow the mass correction for a global motion of the cavity to
differ from zero.

\section*{Partially transmitting mirrors}

In a more satisfactory treatment, perfect mirrors are replaced by partially
transmitting mirrors, described by reflection and transmission amplitudes
respectively denoted $r_{i}$ and $s_{i}$ for the mirror $i=1,2$ obeying
unitarity, causality and high frequency transparency requirements \cite
{Inertia2}. These mirrors are more easily shown to obey causality \cite
{Inertia7}, the divergences associated with the infiniteness of vacuum
energy are regularized \cite{Inertia2,Inertia8}, and the stability problem
arising for a perfect mirror may be solved \cite{Inertia17}. A resonant
enhancement of the motional Casimir force subsists at the optical resonance
frequencies of the Fabry-Perot cavity \cite{Inertia8,Inertia18}.

The susceptibility functions $\chi _{ij}$ may be written (see eqs 20 and 21
of ref. \cite{Inertia8}; $\varepsilon $ is the sign function) 
\[
\chi _{ij}[\omega ]=\frac{i\hbar }{2c^ 2 }\int \frac{{\rm d}\omega ^\prime }
{2\pi }\omega ^\prime (\omega -\omega ^\prime )\varepsilon (\omega
^\prime )\gamma _{ij}^{R}[\omega ^\prime ,\omega -\omega ^\prime ]
\]
where the coefficients $\gamma _{ij}^{R}$ are the sum of two parts 
\[
\gamma _{ij}^{R}[\omega ,\omega ^\prime ]=\gamma _{ij}^{S}[\omega ,\omega
^\prime ]+\gamma _{ij}^{A}[-\omega ,\omega ^\prime ]
\]
Both functions $\gamma _{ij}^{S}$ and $\gamma _{ij}^{A}$ are symmetrical in
the exchange of their two parameters, so that one obtains 
\begin{eqnarray*}
\chi _{ij}[\omega ] &=&\chi _{ij}^{S}[\omega ]+\chi _{ij}^{A}[\omega ] \\
\chi _{ij}^{S}[\omega ] &=&\frac{i\hbar }{2c^ 2 }\int_{0}^{\omega }\frac
{{\rm d}\omega ^\prime }{2\pi }\omega ^\prime (\omega -\omega ^\prime )
\gamma _{ij}^{S}[\omega ^\prime ,\omega -\omega ^\prime ] \\
\chi _{ij}^{A}[\omega ] &=&\frac{i\hbar }{2c^ 2 }\int_{0}^{\infty }\frac
{{\rm d}\omega ^\prime }{2\pi }\omega ^\prime \left( (\omega +\omega
^\prime )\gamma _{ij}^{A}[\omega ^\prime ,\omega +\omega ^{\prime
}]+(\omega -\omega ^\prime )\gamma _{ij}^{A}[-\omega ^\prime ,\omega
-\omega ^\prime ]\right) 
\end{eqnarray*}
The functions $\chi _{ij}^{S}$ scale as $\omega ^{3}$ in the vicinity of
zero frequency as the susceptibility function for a single mirror, and they
do not contribute to the quasistatic coefficients $\kappa _{ij}$, $\lambda
_{ij}$ and $\mu _{ij}$, which we are interested in. We will not discuss them
in more detail. The functions $\gamma _{ij}^{A}$ are given by 
\begin{eqnarray*}
\gamma _{11}^{A}[\omega ,\omega ^\prime ] &=&\frac{\left( r_1 [\omega
]+r_1 [\omega ^\prime ]\right) \left( r_2 [\omega ]e^{2i\omega \tau
}+r_2 [\omega ^\prime ]e^{2i\omega ^\prime \tau }\right) }{d[\omega
]d[\omega ^\prime ]} \\
\gamma _{21}^{A}[\omega ,\omega ^\prime ] &=&-\frac{\left( r_1 [\omega
]+r_1 [\omega ^\prime ]\right) \left( r_2 [\omega ]+r_2 [\omega
^\prime ]\right) e^{i(\omega +\omega ^\prime )\tau }}{d[\omega ]d[\omega
^\prime ]} \\
d[\omega ] &=&1-r_1 [\omega ]r_2 [\omega ]e^{2i\omega \tau }
\end{eqnarray*}
$\gamma _{22}^{A}$ and $\gamma _{12}^{A}=\gamma _{21}^{A}$ are obtained by
exchanging the roles of the two mirrors. Straightforward differentiations of
the functions $\chi _{ij}^{A}$ then lead to the quasistatic coefficients
(see eqs 5) 
\begin{eqnarray*}
\kappa _{ij} &=&-\frac{i\hbar }{2c^ 2 }\int_{0}^{\infty }\frac{{\rm d}\omega 
}{2\pi }\omega ^ 2 \left( \gamma _{ij}^{A}[\omega ,\omega ]-\gamma
_{ij}^{A}[-\omega ,-\omega ]\right)  \\
\lambda _{ij} &=&\frac{\hbar }{2c^ 2 }\int_{0}^{\infty }\frac{{\rm d}\omega 
}{2\pi }\omega \left( \gamma _{ij}^{A}[\omega ,\omega ]+\gamma
_{ij}^{A}[-\omega ,-\omega ]+\omega \gamma _{ij}^{A}{}^\prime [\omega
,\omega ]-\omega \gamma _{ij}^{A}{}^\prime [-\omega ,-\omega ]\right)  \\
\mu _{ij} &=&\frac{i\hbar }{4c^ 2 }\int_{0}^{\infty }\frac{{\rm d}\omega }
{2\pi }\omega \left( 2\gamma _{ij}^{A}{}^\prime [\omega ,\omega ]+2\gamma
_{ij}^{A}{}^\prime [-\omega ,-\omega ]+\omega \gamma _{ij}^{A}{}^{\prime
\prime }[\omega ,\omega ]-\omega \gamma _{ij}^{A}{}^{\prime \prime }[-\omega
,-\omega ]\right) 
\end{eqnarray*}
It is understood that differentiation only bears on one of the two frequency
parameters 
\begin{eqnarray*}
\gamma _{ij}^{A}{}^\prime [\omega ,\omega ] &=&\left( \partial _{\omega
}\gamma _{ij}^{A}[\omega ,\omega ^\prime ]\right) _{\omega ^{\prime
}=\omega }=\frac 1  2 \frac{{\rm d}\gamma _{ij}^{A}[\omega ,\omega ]}{{\rm d}
\omega } \\
\gamma _{ij}^{A}{}^{\prime \prime }[\omega ,\omega ] &=&\left( \partial
_{\omega }^ 2 \gamma _{ij}^{A}[\omega ,\omega ^\prime ]\right) _{\omega
^\prime =\omega }
\end{eqnarray*}

One checks \cite{Inertia8} that the static coefficients $\kappa _{ij}$ fit
the variation of the mean Casimir force $F$ between the two partially
transmitting mirrors 
\begin{eqnarray}
&&\kappa _{11}=\kappa _{22}=-\kappa _{12}=-\kappa _{21}=\frac{{\rm d}F}
{{\rm d}q}  \eqnum{7a} \\
&&F=\frac{\hbar }{c}\int_{0}^{\infty }\frac{{\rm d}\omega }{2\pi }\omega
\left( 1-\frac 1 {d[\omega ]}+1-\frac 1 {d[-\omega ]}\right)   \eqnum{7b}
\end{eqnarray}
The viscosity coefficients $\lambda _{ij}$ remain equal to zero for
partially transmitting mirrors 
\begin{equation}
\lambda _{ij}=0  \eqnum{8}
\end{equation}
One eventually rewrites the inertia corrections 
\begin{eqnarray}
\mu _{ij} &=&\frac{i\hbar }{4c^ 2 }\int_{0}^{\infty }\frac{{\rm d}\omega }
{2\pi }\omega ^ 2 \left( \Gamma _{ij}[\omega ]-\Gamma _{ij}[-\omega ]\right) 
\eqnum{9a} \\
\Gamma _{ij}[\omega ] &=&\gamma _{ij}^{A}{}^{\prime \prime }[\omega ,\omega
]-\frac{{\rm d}\gamma _{ij}^{A}{}^\prime [\omega ,\omega ]}{{\rm d}\omega }
=-\left( \partial _{\omega }\partial _{\omega ^\prime }\gamma
_{ij}^{A}[\omega ,\omega ^\prime ]\right) _{\omega ^\prime =\omega } 
\nonumber \\
\Gamma _{11}[\omega ] &=&-2\frac{r_1 ^\prime [\omega ]e^{2i\omega \tau
}\left( 2i\tau r_2 [\omega ]+r_2 ^\prime [\omega ]\right) }{d[\omega
]^ 2 }-4\frac{d^\prime [\omega ]^ 2 }{d[\omega ]^{4}}  \eqnum{9b} \\
\Gamma _{21}[\omega ] &=&-4i\tau \frac{d^\prime [\omega ]}{d[\omega ]^ 2 }
+4\tau ^ 2 \frac{1-d[\omega ]}{d[\omega ]^ 2 }+2\frac{r_1 ^\prime [\omega
]r_2 ^\prime [\omega ]e^{2i\omega \tau }}{d[\omega ]^ 2 }+4\frac
{d^\prime [\omega ]^ 2 }{d[\omega ]^{4}}  \eqnum{9c}
\end{eqnarray}
$\Gamma _{22}$ and $\Gamma _{12}=\Gamma _{21}$ are obtained by exchanging
the roles of the two mirrors. At the limit of perfect reflection, the mass
corrections (6) are recovered. They are proportional to $\frac{\hbar }{cq}$
in this limit, as it could have been guessed from a dimensional analysis
since there are no other dimensioned parameters than $\hbar $, $c$ and $q$.
For partially transmitting mirrors, the mass corrections are no longer
homogeneous functions of the distance $q$ between the two mirrors, since
they now depend upon the reflectivity functions and particularly upon the
reflection cutoff frequencies.

\section*{Mass correction for the compound system}

We come to the study of a global motion of the compound system, when the two
mirrors move with the same acceleration 
\[
\delta q_1 (t)=\delta q_2 (t)=\delta q(t)
\]
In other words, their distance remains equal to its initial value 
\[
q_2 (t)-q_1 (t)=q_2 (0)-q_1 (0)=q
\]
We notice that the motional force is computed in a first order expansion in
the mirrors' displacement, performed in the vicinity of a static
configuration (mirrors at rest). In particular, Lorentz contraction, which
scales as $\frac{v^ 2 }{c^ 2 }$ with $v$ the velocity of cavity and $c$ the
velocity of light, can be disregarded.

The global force exerted upon the cavity is the sum of the forces exerted
upon the two mirrors 
\[
\delta F(t)=\delta F_1 (t)+\delta F_2 (t)
\]
and the motion of the system is described by the linear susceptibility $\chi 
$ associated with the total force $F$ 
\[
\delta F[\omega ]=\chi [\omega ]\delta q[\omega ]\qquad \chi [\omega
]=\sum_{i}\sum_{j}\chi _{ij}[\omega ]
\]
The quasistatic expansion (5) now becomes 
\begin{eqnarray*}
&&\delta F(t)=-\left( \kappa \delta q(t)+\lambda \delta q^\prime (t)+\mu
\delta q^{\prime \prime }(t)+\ldots \right)  \\
&&\kappa =\sum_{i}\sum_{j}\kappa _{ij}\qquad \lambda
=\sum_{i}\sum_{j}\lambda _{ij}\qquad \mu =\sum_{i}\sum_{j}\mu _{ij}
\end{eqnarray*}
The static coefficient $\kappa $ vanishes (see eqs 7), as required by
invariance in a global translation of the compound system, or equivalently,
by the fact that Casimir force only depends upon the distance between the
two mirrors. The viscosity coefficient $\lambda $ also vanishes (see eqs 8),
in consistency with Lorentz invariance of vacuum. The force computed for a
global motion of the cavity is eventually an inertial force at the
quasistatic limit, where the higher order terms are negligible when compared
to the second order one 
\begin{equation}
\delta F(t)=-\mu \delta q"(t)\qquad \mu =\sum_{i}\sum_{j}\mu _{ij} 
\eqnum{10}
\end{equation}
This relation is exact in the particular case of a uniform acceleration
where the higher order terms do vanish.

In the limiting case of perfect mirrors, we obtain from equations (4) 
\begin{eqnarray}
\chi [\omega ] &=&\widetilde{\xi }[\omega ]+i\xi [\omega ]  \eqnum{11a} \\
\xi [\omega ] &=&\frac{\hbar }{6\pi c^ 2 }\omega ^{3}  \eqnum{11b} \\
\widetilde{\xi }[\omega ] &=&\frac{\hbar }{6\pi c^ 2 }\left( \omega
^{3}-\omega \frac{\pi ^ 2 }{\tau ^ 2 }\right) \tan \frac{\omega \tau } 2  
\eqnum{11c}
\end{eqnarray}
The dissipative part $\xi $ of the susceptibility is the same for the
compound system as for a single perfect mirror. It follows from the
fluctuations-dissipation relations that fluctuations of the global force are
also the same as for a single mirror. These properties mean that the
compound system may, at least for dissipation and fluctuations, be
considered as an individual object. They correspond to the fact that the
dissipative functions $\xi _{ij}$ coincide precisely in the case of perfect
mirrors (see the discussion following eqs 4) with the contributions of outer
space.

In contrast, the dispersive part $\widetilde{\xi }$ of the susceptibility
differs from the single mirror case. In particular, it contains a mass
correction, whereas such a correction was zero for a single mirror 
\[
\mu =\frac{\chi ^{\prime \prime }[0]} 2 =-\frac{\hbar \pi }{12cq}
\]
This means that the field energy stocked inside the cavity, and bound along
its motion, contributes to its inertia. Furthermore, the mass correction
appears to be directly connected to the Casimir energy $U$ given by equation
(3) 
\begin{equation}
\mu c^ 2 =2U  \eqnum{12}
\end{equation}
This relation explains the negative sign of the mass correction, since
Casimir energy is a binding energy. However, the factor 2 seems to prevent a
simple explanation of the mass correction from the law of inertia of energy 
\cite{Inertia3}. A precise discussion requires a more detailed analysis of
the law of inertia of energy, and is delayed to the next section.

It is worth recalling that the mass correction $\mu $ has been calculated at
the quasistatic limit. The following relation, deduced from equations (11) 
\[
\widetilde{\xi }[\omega ]=\mu \omega ^ 2 \left( 1-\left( \frac{\omega \tau }
{\pi }\right) ^ 2 \right) \frac{\tan \frac{\omega \tau } 2 }{\frac{\omega
\tau } 2 }
\]
shows that the motional force behaves as an inertial force at low
frequencies only ($\omega \tau \ll 1$ that is $\omega q\ll c$). The
dispersive part $\widetilde{\xi }[\omega ]$ of the susceptibility presents
divergences at the frequencies $\omega =m\frac{\pi }{\tau }$ ($m$ an odd
integer except $m=\pm 1$) which result from a constructive interference
between the different numbers of cavity roundtrips \cite{Inertia8} and may
in principle be observable for an arbitrarily small magnitude of the
frequency component $\delta q[\omega ]$, that is for a velocity of the
system remaining much smaller than the velocity of light. In other words, it
turns out that the internal optical modes of the Fabry-Perot cavity are
coupled to its external mechanical motion, even for a global motion where
the distance between the mirrors is constant. At the adiabatic limit, this
coupling may be considered as responsible for an inertia correction. At
higher frequencies, internal resonances of the Fabry-Perot cavity are
efficiently excited and the effective inertia may become large.

For a cavity built with two partially transmitting mirrors, $\mu $ is
deduced (see eq. 10) from equations (9) 
\begin{eqnarray*}
&&\mu =\frac{i\hbar }{4c^ 2 }\int_{0}^{\infty }\frac{{\rm d}\omega }{2\pi }
\omega ^ 2 \left( \Gamma [\omega ]-\Gamma [-\omega ]\right)  \\
&&\Gamma [\omega ]=\sum_{i}\sum_{j}\Gamma _{ij}[\omega ]=-4i\tau \frac
{d^\prime [\omega ]}{d[\omega ]^ 2 }
\end{eqnarray*}
This relation can be transformed by an integration by parts into an
expression in terms of the mean Casimir force given by equations (7) 
\begin{equation}
\mu =-\frac{2Fq}{c^ 2 }  \eqnum{13a}
\end{equation}
Expression (12) is recovered at the limit of perfect mirrors, since $(-Fq)$
then coincides with the Casimir energy $U$ ($U$ scales as $\frac 1 {q}$ and 
$F=\frac{{\rm d}U}{{\rm d}q}$). In general, the force $F$ is the difference
between the mean energy densities inside and outside the cavity, and the
quantity $(-Fq)$ is the integral $E_{f}$ of the field energy density,
measured as a difference with respect to the mean density in free space ($e_
{\rm inner}$ and $e_{\rm outer}$ are the energy densities inside and
outside the cavity \cite{Inertia2}) 
\begin{equation}
-Fq=\left( e_{\rm inner}-e_{\rm outer}\right) q=E_{f}  \eqnum{13b}
\end{equation}
Causal reflectivity functions fulfilling the high frequency transparency
requirements have to be frequency dependent. Then, the field experiences
reflection delays upon each mirror, and the Casimir energy $U$ may be
written as the sum of the integrated field energy $E_{f}$ and of an extra
contribution, attributed to an apparent modification of the cavity length
associated with reflection delays (see eq. 27 in ref. \cite{Inertia2}). This
extra contribution is much smaller than $E_{f}$ at the limit of short
delays, when $\tau $ is greater than the reflexion delays.

For partially transmitting mirrors, the divergences of the motional
susceptibilty at the optical resonance frequencies $\omega =m\frac{\pi }
{\tau }$ ($m$ an odd integer except $m=\pm 1$) of the cavity are regularized
and become dispersion-shaped resonances \cite{Inertia8}.

\section*{Inertia of a stressed rigid body}

The mass corrections obtained in the foregoing section differ from the
expression $\frac{E_{f}}{c^ 2 }$ which would naively be expected from the
law of inertia of energy \cite{Inertia3}. It is worth refering this problem
to arguments given by Einstein in his first survey article on relativity 
\cite{Inertia14}.

The point is that the cavity moves as a rigid body, the two mirrors having
the same acceleration, while being submitted to a force. In this situation,
the momentum $P$ of the cavity has to be written \cite{Inertia19} 
\begin{equation}
P=\left( m+\delta m\right) v  \eqnum{14a}
\end{equation}
where $v$ is the global velocity $v=q_1 ^\prime =q_2 ^\prime $, $m$
the genuine mass of the mirrors, and $\delta m$ a mass correction which
depends upon the field energy $E_{f}$ and the force $F$ 
\begin{equation}
\delta m=\frac{E_{f}-Fq}{c^ 2 }  \eqnum{14b}
\end{equation}
The global force exerted upon the system is therefore 
\begin{equation}
\frac{{\rm d}P}{{\rm d}t}=\left( m+\delta m\right) a  \eqnum{14c}
\end{equation}
where $m$ and $\delta m$ are considered as constant and $a=v^\prime $ is
the global acceleration. It thus follows from relativistic considerations
(and not from the particular dependence of $F$ or $E_{f}$ versus $q$) that
inertia of a stressed rigid body is not given by the simple expression 
$\frac{E_{f}}{c^ 2 }$, but by the more elaborate one (14) which involves the
value of the stress. This refinement plays a role in the relativistic
analysis of a thermodynamic system with a homogeneous normal pressure \cite
{Inertia20,Inertia21}.

For the problem studied in the present paper, the stress is the Casimir
force and the quantities $E_{f}$ and $(-Fq)$ coincide, so that the mass
given by equations (13) and computed from the motional susceptibility fits
expressions (14). This proves not only that Casimir energy does contribute
to inertia of the cavity, but that its contribution is precisely what is
expected from the law of inertia of energy.

It has however to be emphasized that the usual expression $\frac{E_{f}}{c^ 2}
$ effectively describes the inertia correction associated with Casimir
energy, when the motion of the relativistic center of inertia of the whole
system (mirrors and stocked fields) is considered. This follows directly
from the first statement of the law of inertia of energy \cite{Inertia3},
which is of course equivalent to the second statement corresponding to
equations (14), as it can be checked by defining the energy $E$, the
momentum $P$, and the relativistic center of inertia $Q$ of the compound
system 
\begin{eqnarray}
E &=&e_1 +e_2 +E_{f}\qquad P=p_1 +p_2 +P_{f}  \nonumber \\
&&EQ=e_1 q_1 +e_2 q_2 +E_{f}\frac{q_1 +q_2 } 2   \eqnum{15a}
\end{eqnarray}
where $e_{i}$ and $p_{i}$ are the relativistic energy and momentum of the
mirror $i=1,2$; $E_{f}$ and $P_{f}$ are the energy and momentum of the
stocked field; we have used the fact that the stocked field energy is
distributed homogeneously inside the cavity and has therefore its center of
inertia at the middle point $\frac{q_1 +q_2 } 2 $. Computing explicitly
the time derivative of the center of inertia $Q$ defined above and noting
that 
\[
p_{i}=e_{i}\frac{q_{i}^\prime }{c^ 2 }\qquad e_{i}^\prime =p_{i}^{\prime
}q_{i}^\prime =F_{i}q_{i}^\prime \qquad E^\prime =0
\]
one shows that equations (14) are equivalent to 
\begin{equation}
c^ 2 P=e_1 q_1 ^\prime +e_2 q_2 ^\prime +\left( E_{f}-Fq\right) 
\frac{q_1 ^\prime +q_2 ^\prime } 2 =EQ^\prime   \eqnum{15b}
\end{equation}
The center of inertia $Q$ thus behaves as the position of a particle of mass 
$\frac{E}{c^ 2 }$.

\section*{Discussion}

The main result obtained in the present paper is that Casimir energy, that
is a change in vacuum energy, does contribute to the inertial mass of a
Fabry-Perot cavity. The computed mass agrees with the prediction of the law
of inertia of energy, when the fact that the cavity moves as a stressed
rigid body is accounted for. The equivalence principle then tells us that
Casimir energy has also to contribute to gravity.

The calculations have been performed in the simple case of a scalar field in
a 2D spacetime, and they would have to be generalized to scatterers and
vacuum fields in a 4D spacetime. In their present form however, they already
meet interesting questions concerning the nature of inertia.

Einstein suggestively stated \cite{Inertia3} that {\it ``radiation conveys
inertia between emitting and absorbing bodies''}. In the context of the
present paper, it must be understood that vacuum fields stocked inside the
cavity convey inertia between the two mirrors. The stocked energy is bound
to intracavity space, between the two mirrors, rather than to the mirrors
themselves (this appears clearly in eqs 15). This is why its contribution to
inertia gives rise to properties usually associated with the ``principle of
relativity of inertia''.

Following ideas expressed by Mach, Einstein \cite{Inertia22} attempted to
define such a principle while he was developing his theory of general
relativity. Later on \cite{Inertia23}, he described as follows the
properties associated with such a conception: (1) {\it The inertia of a body
must increase when ponderable masses are piled up in his neighborhood}; (2) 
{\it A body must experience an accelerating force when neighboring masses
are accelerated;} a third property, concerning rotation, can be disregarded
in comparing with calculations in a 2D spacetime. Such effects are actually
predicted by general relativity, with a very small magnitude however \cite
{Inertia24}.

Although they are derived from a theoretical study of vacuum fluctuations
rather than from a study of gravity, the results of the present paper partly
fit these requirements. Indeed, the mass of a scatterer in vacuum is
modified by the presence of another scatterer, and the correction depends
upon the relative distance $q$ of the two scatterers. Each scatterer
experiences a force proportional to the acceleration of the other one. Note
that a global acceleration of the whole system gives rise to a force, in
agreement with the law of inertia of energy.

As a consequence, it appears fruitful to consider the vacuum, the quantum
``empty space'', as a Lorentz-invariant realization of inertial reference
frame \cite{Inertia25}, the inertial forces representing the reaction of
vacuum fields to an accelerated motion with respect to them. The results of
the present paper make clear that this conception is pertinent for those
inertial forces which are associated with a stocked field energy like the
Casimir energy. An appealing feature of this conception is that it would
make plausible that gravity forces are actually equivalent to a modification
of vacuum fields \cite{Inertia26}. It has nevertheless to be acknowledged
that difficulties plague the possibility of explaining all inertia along
these lines.

It appears that the effects discussed in the present paper depend on the
scattering properties of neighboring scatterers, while no direct relation
has been established between these scattering properties and the masses.
Furthermore, the magnitude and sign of the mass corrections do not fit
Einstein's requirements. Negative mass corrections are obtained, as expected
from the fact that Casimir energy is a binding energy. Then, the mass
corrections obtained in the present paper scale as $\frac{\hbar }{cq}$ for a
perfect mirror; they are negligible when compared to the mirror's mass $m$,
as soon as the distance $q$ is greater than the Compton wavelength $\frac
\hbar {mc}$. It can nevertheless be noted that, if the magnitude of the
effect is the same in a 4D spacetime (this would be consistent with
dimensional analysis), the mass corrections may become large and even
diverge, when scatterers uniformly distributed in space are considered. The
effect of the finite time of flight between bodies has to be kept in mind,
since distant scatterers can only modify the inertial force with large time
delays.

\medskip \noindent {\bf Acknowledgements}

Thanks are due to J.-M. Courty, A. Heidmann and P.A. Maia Neto for
discussions.


\begin{references}
\bibitem{Inertia1}  Casimir H.B.G., {\it Proc. K. Ned. Akad. Wet.} {\bf 51}
793 (1948); a recent review may be found in: Plunien G., M\"{u}ller B. and
Greiner W., {\it Phys. Rep.} {\bf 134} 87 (1986).

\bibitem{Inertia2}  Jaekel M.T. and Reynaud S., {\it J. Physique} {\bf I 1}
1395 (1991).

\bibitem{Inertia3}  Einstein A., {\it Ann. Physik} {\bf 18} 639 (1905)
[reprinted in english in {\it The Principle of Relativity} (Dover
Publications, 1952)]; {\it Ann. Physik} {\bf 20} 627 (1906).

\bibitem{Inertia4}  Pais' biography of Einstein gives an historical account
of the successive demonstrations of this law: Pais A. {\it Subtle is the
Lord...} (Oxford University Press, 1982), ch.7.

\bibitem{Inertia5}  See for instance: Feynman R.P. and Hibbs A.R., {\it 
Quantum Mechanics and Path Integrals} (Mac Graw Hill, 1965), p.244; Enz
C.P., in {\it Physical Reality and Mathematical Description} C.P.Enz and
J.Mehra eds, (Dordrecht, 1974) p.124; a recent discussion containing
references is given in: Wesson P.S., {\it Astrophys. J.} {\bf 378} 466
(1991).

\bibitem{Inertia6}  Sciama D.W., in {\it The Philosophy of Vacuum}
S.Saunders and H.R.Brown eds, (Clarendon, 1991), p.137.

\bibitem{Inertia7}  Jaekel M.T. and Reynaud S., {\it Quant. Opt.} {\bf 4} 39
(1992).

\bibitem{Inertia8}  Jaekel M.T. and Reynaud S., {\it J. Physique} {\bf I 2}
149 (1992).

\bibitem{Inertia9}  Barton G., {\it J. Phys.} {\bf A24} 991 (1991); {\it J.
Phys.} {\bf A24} 5533 (1991); in {\it Cavity Quantum Electrodynamics}
(Suppl. Adv. Atom. Mol. Opt. Phys.), P.Berman ed., (Academic Press, 1994).

\bibitem{Inertia10}  Fulling S.A. and Davies P.C.W., {\it Proc. R. Soc.
London} {\bf A348} 393 (1976).

\bibitem{Inertia11}  Boyer T.H., {\it Sci. Am.} {\bf 253} 56 (1985).

\bibitem{Inertia12}  Hawking S.W., {\it Commun. Math. Phys.} {\bf 43} 199
(1975); Davies P.C.W., {\it J. Phys.} {\bf A8} 609 (1975); Unruh W.G., {\it 
Phys. Rev.} {\bf D14} 870 (1976); Birrell N.D. and Davies P.C.W., {\it 
Quantum fields in curved space} (Cambridge, 1982), and references therein.

\bibitem{Inertia13}  Note however that friction and mass corrections appear
for a mirror in thermal fields: Jaekel M.T. and Reynaud S., {\it Phys. Lett.}
{\bf A172} 319 (1993).

\bibitem{Inertia14}  Einstein A., {\it Jahrb. Radioakt. Elektron.} {\bf 4}
411 (1907), {\bf 5} 98 (1908) [translated in english and commented by
Schwartz H.M., {\it Am. J. Phys.} {\bf 45} 512, 811, 899 (1977)].

\bibitem{Inertia15}  We have added the contributions of intracavity and
outer fields, both contributions being derived from the results of Fulling
and Davies \cite{Inertia10}; the same expression is obtained as the limit of
perfect reflection in ref. \cite{Inertia7}.

\bibitem{Inertia16}  Landau L.D. and Lifshitz E.M. {\it Cours de Physique
Th\'{e}orique: Physique Statistique} (Mir, 1967) ch.12; Kubo R., {\it Rep.
Progr. Phys.} {\bf 29} 255 (1966).

\bibitem{Inertia17}  Jaekel M.T. and Reynaud S., {\it Phys. Lett.} {\bf A167}
227 (1992).

\bibitem{Inertia18}  Resonant enhancement of the interaction of atoms with
vacuum has been studied for instance by: Kleppner D., {\it Phys. Rev. Lett.} 
{\bf 47} 233 (1981); Haroche S., in {\it New Trends in Atomic Physics}
G.Grynberg and R.Stora eds (North Holland, Amsterdam, 1984), p.193;
enhancement of vacuum radiation pressure in a cavity has also been studied
by: Braginski V.B. and Khalili F.Ya., {\it Phys. Lett.} {\bf A161} 197
(1991).

\bibitem{Inertia19}  We rewrite eqs (18c) and (19) of ref. \cite{Inertia14}
with the appropriate changes of notation: $v$ stands for the global velocity
of the system (in place of $q$ in ref. \cite{Inertia14}); corrections of the
order of $\frac{v^ 2 }{c^ 2 }$ are neglected; $E_{f}$ stands for the
internal energy ($E_{0}$), $F=F_1 =-F_2 $ for the internal force ($K_{0}$), 
$q$ for the distance between the two points of application of the force 
($\delta _{0}$).

\bibitem{Inertia20}  In equation (14.b), $(E_{f}-Fq)$ is replaced by 
$(E_{f}+pV)$, where $p$ is the pressure and $V$ the volume (see ref. \cite
{Inertia14}); the change of sign is due to the fact that a positive pressure 
$p$ represents a repulsion between the two points of application, while a
positive force $F$ is defined here as an attraction.

\bibitem{Inertia21}  The related case of mechanical systems containing
stressed threads or rods, as well as controversies held on it since the
birth of relativity up to recent times, are presented for instance in:
Martins R. de A., {\it Am. J. Phys.} {\bf 50} 1008 (1982).

\bibitem{Inertia22}  Einstein A., {\it Vierteljahrsschrift f. Gerichtliche
Medizin.} {\bf 44} 37 (1912); this attempt is described for example by:
Kastler A., {\it M\'{e}moires de la Classe des Sciences de l'Acad\'{e}mie
Royale de Belgique} 2e s\'{e}rie {\bf 44} (1) 13 (1981) [reprinted in {\it 
Oeuvre Scientifique} (Editions du CNRS, Paris 1988) p.1230]; Pais A. \cite
{Inertia4} ch.15e.

\bibitem{Inertia23}  Einstein A., {\it The Meaning of Relativity} (Princeton
University Press, 1946).

\bibitem{Inertia24}  Rosen N., in {\it To fulfill a vision} Y.Ne'eman ed.,
(Addison Wesley, 1981), ch.5.

\bibitem{Inertia25}  This idea has often been expressed; see as an example:
De Witt B.S., in {\it General relativity: An Einstein Centenary Survey}
S.W.Hawking and W.Israel eds, (Cambridge, 1979), ch.14.

\bibitem{Inertia26}  See for example: Dicke R.H., {\it Rev. Mod. Phys.} {\bf 
29} 363 (1957); Sakharov A.D., {\it Doklady Akad. Nauk} {\bf 177} 70 (1967) 
[{\it Sov. Phys. Doklady} {\bf 12} 1040 (1968)]; see also a list of
references in: Puthoff H.E., {\it Phys. Rev.} {\bf A39} 2333 (1989).
\end{references}
\end{document}